# On the initiation of fiber fuse damage in high-power ytterbium-doped fiber lasers

JIADING TIAN,[1,2] ZEHUI WANG,[3] QIRONG XIAO,[1,2,*] DAN LI,[1,2] PING YAN,[1,2] AND MALI GONG[1,2]

[1]*Ministry of Education Key Laboratory of Photonic Control Technology, Department of Precision Instrument, Tsinghua University, Beijing 100084, China*
[2]*State Key Laboratory of Precision Measurement Technology and Instruments, Department of Precision Instrument, Tsinghua University, Beijing 100084, China*
[3]*Department of Laser Equipment, Jiangsu Shuguang Photoelectric Co. Ltd.,Yangzhou 225009, China*
*\*xiaoqirong@mail.tsinghua.edu.cn*

**Abstract:** Fiber fuse effect can occur spontaneously and propagate along optical fibers to cause wide-spread damage; it threatens all applications involving optical fibers. This paper presents two results. First, it establishes that the initiation of fiber fuse (IFF) in silica fibers is caused by defect-induced absorption. Critical temperatures and critical optical powers for IFF are simulated for the first time using a 3D solid-state heat transfer model with heat source generated by defect-induced absorption. In this method, formation energies of the defects can be uniquely determined, which offers critical information on the chemical reasons for fiber fuse. Second, this paper offers a method to evaluate operating temperatures of fiber lasers. General analytical solutions of the operating temperatures along gain fibers are deduced. Results of 976-nm laser-diode-(LD)-pumped and 1018-nm tandem-pumped ytterbium-doped fiber (YDF) amplifiers using 10/130-µm YDFs are calculated. Potential limits caused by fiber fuse are discussed.

## 1. Introduction

Optical fibers are ubiquitous materials that are used not only for delivering light, but also for generating light via different kinds of fiber lasers. They play important roles in both scientific research and various applications such as mass manufacturing, energy, and medical treatment. However, as the optical powers that optical fibers handle keep increasing, spontaneous damage of the optical fibers is surging nowadays. Spontaneous damage of silica fibers can evolve into several types of phenomena. The fiber fuse effect is the most spectacular type, in which damage propagates along fiber in typically ~m/s speed with bright, visible light emission. Camera video frames showing the propagation of fiber fuse are given in Fig. 1. The measured speed of the propagation of fiber fuse (PFF) under high average optical power can be >14 m/s [1]. It means catastrophic destruction of a whole optical system can be instantaneous. Spontaneous fiber fuse is a serious threat to optical systems connecting to optical fibers. In fact, it is widely witnessed in both high-power continuous-wave fiber lasers and ultrafast fiber lasers, usually in those of kilowatt-level-or-higher optical powers or microjoule-level-or-higher pulse energies. However, as fiber fuses are negative experimental results (like other unwanted damages), they are mostly dealt with in the middle of research instead of appearing as documented results. Even if they appear in reports, in most cases only 'thermal damage' [2-7] or other general descriptions are mentioned instead of 'fiber fuse'. Thus, the severity of spontaneous fiber fuse damage may be substantially undervalued in the literature. Nevertheless, there are still a surge of reports in recent years directly on characteristics of the PFF in various kinds of optical fibers [8-11], including photonic bandgap fibers [12], hollow-core fibers [13] and polymer fibers [14]. The name 'fiber fuse' is explicitly discussed with a rising frequency in many applications, e.g., erbium-doped fiber lasers [15, 16] and quantum encrypted communications [17]. Besides, the usage of fiber fuse is quickly extending to glass drilling [18, 19], fabrication of spherical in-fiber microcavity [20] and demonstration of various kinds of fiber sensors [14, 21-25].

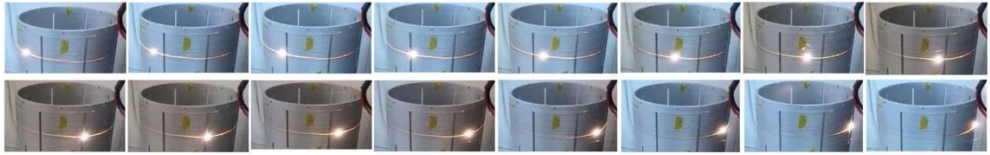

Fig. 1. Continuous frames (30 fps) of a camera video of the propagation of fiber fuse (PFF). The cylinder on which the fiber is coiled has a diameter of ~50 cm.

In the state-of-the-art techniques, blocking PFF is possible only when in-fiber optical power is watt-level [26, 27]. In fiber lasers and systems operating with higher optical powers, fiber fuses run with basically no hinderance and cause huge losses. For controlling the risk, studying the reasons why fiber fuses spontaneously happen and, particularly, under what quantitative conditions they happen, are highly important. However, due to many limitations in experimental capabilities of observing the damage effect closely, as well as high cost of such experiments, results to date only unveil small pieces of the puzzle. It is only empirically and qualitatively known that high thermal loads and nonlinear optical effects, such as stimulated Raman scattering (SRS), are harmful [28]. Previous theories on fiber fuse offer little quantitative information about the initiation of fiber fuse (IFF); this will be elaborated in Section 2.1. In our previous experimental studies [29], critical temperatures and critical optical powers for IFF in 10/130-µm ytterbium-doped fibers (YDFs) were measured for the first time. Results unveiled a special correlation between the critical temperature and the critical optical power for IFF, which led to a parameter $u_0$ of the dimension of energy. It was interpreted that the parameter relates to specific kinds of chemical processes, but with no further proof then. There are still crucial gaps between the experimental result and such physical interpretations, which need a tighter stitching; this problem will be elaborated in Section 2.2.

This paper deals with fiber fuse in YDFs and it presents majorly two results. First, it offers a theoretical investigation of the physical mechanism of IFF. A 3D solid-state heat transfer model with heat source from defect-induced absorption of laser light is built. By finding suitable formation energies of defects, it is found that the simulation results of critical temperatures and critical optical powers for IFF can perfectly match our previous experimental results of the critical conditions for IFF. The method can uniquely determine the formation energies of defects $u_F$ and absorption coefficients $\alpha_0$ that the defects cause, both of which are important intrinsic properties of the fibers. It finds that $u_F$ is not, in fact, but is larger than, the previously experimentally revealed parameter $u_0$. The results technically prove that a defect-induced absorption is the direct cause for spontaneous IFF in silica optical fibers. Second, this paper offers a method to evaluate operating temperatures of fiber lasers. General analytical solutions of the operating temperatures are deduced. Results of 976-nm laser-diode-(LD)-pumped and 1018-nm tandem-pumped amplifiers using 10/130-µm YDFs are calculated. The results show that the critical conditions for IFF may not be met in most cases of state-of-the-art experimental setups if deterioration and burning of polymer coatings of the fibers can be avoided. The results suggest a large space for increasing the output powers of YDF lasers. The method provides a quantitative guidance for future design of high-power fiber lasers.

## 2. Experimental results of IFF

### 2.1 Assumptions from previous studies

To establish this work, it is necessary to review some previous results. In fact, there can still be sharp and fundamental divisions of the understanding on how fiber fuses happen spontaneously, among people of different backgrounds of knowledge and experiences. For researchers familiar with damages of bulk optics, silica fiber fuse can be ascribed to one of the following two reasons:

(1) intrinsic damage that optical power density of laser light exceeds the laser-induced damage threshold (LIDT) of the silica materials, (2) external factors, such as surface defects and contaminations. However, a prominent difference is that, fiber fuse can propagate in silica fibers with very low optical power density (usually ~MW/cm$^2$), which is 3-order lower than the LIDT of silica (~GW/cm$^2$). This has caused a long-time confusion that basically separate the studies of the above two topics and seems to require a new theory for fiber fuse. Nevertheless, fiber fuses show many features of thermal damage, such as its characteristic trace of bullet-shaped in-fiber cavities that are likely related to fluid instability of melted fiber material under high temperature. It is possible that the description of its physics can be based on the mechanism of its heat generation. On this issue, there are many hypotheses that can be divided into the following two types (I) and (II) based on their aspects of description.

(I) The heating is caused by certain external changes from the original ideal, intact state of optical fibers, e.g., cracks or contaminations. In practice, spontaneous fiber fuses do sometimes initiate at the end facets of fibers. Therefore, it is convenient to assume that fiber fuses will not exist as long as fibers are kept clean, free of surface contaminations. However, as various kinds of fiber lasers develop fast, more experiences come as fiber fuses can spontaneously happen in the middle of intact silica fibers. An opinion ascribes this to accidental contaminations inside the fibers, e.g., at the interface between the silica cladding and the polymer coating, or structural imperfections (cracks or stresses) caused in manufacturing processes. These cases do exist; in fact, to avoid such cases is one of the design purposes of triple-cladding fibers [4]. However, they alone can explain only some but not all; for example, they alone cannot explain that spontaneous IFFs favorite certain places in fiber lasers, e.g., tens of centimeters after the pump-input fusion-splice point in gain fiber. An explanation can be that the in-core propagating laser light has somehow leaked out of the fiber and heated the surrounding polymer coating [5]; structural defects like cracks can be generated by somehow existing random pulses (e.g., from SRS) that exceed LIDTs, which then scatter out the in-core laser light to cause the heating; or, the cracks themselves can 'directly' absorb and heat up. However, in many cases, it is still hard to verify the existence of such hypothesized pulses or cracks before IFFs, or to explain it. In all, it is a possibility that contaminations or structural imperfections are not necessarily prerequisite for IFF; there can be more fundamental reasons behind.

(II) In a more fundamental perspective, the heating is due to the fiber materials' intrinsic absorption, which generally has a positive correlation with temperature. A high local temperature somehow existing can let absorption rise drastically, which overrides heat dissipation and causes a positive feedback loop to the temperature and the absorption, which then eventually causes the IFF. This theory has been adopted by previous studies on fiber fuse. Hand et al.[30] proposed that there is an absorption $\alpha(T)$ in Ge-doped silica fibers (before the following physical changes of fiber fuse, i.e., when the fiber material is still in solid state) that takes the following form

$$\alpha(T) = \alpha_0 \exp(-E_f/k_B T), \tag{1}$$

where $T$ is temperature, $k_B$ is Boltzmann constant, $\alpha_0$ is a coefficient and $E_f$ is believed to be the activation energy of Ge-related chemical defects that causes the absorption. The value of $E_f$ took 2.2 eV and $\alpha_0$ took 1.2×10$^{-6}$ m$^{-1}$ for fitting the measured absorption data. Behind Eq. 1 an implicit presumption is that the absorption is linearly dependent on the concentration of chemical defects ($\propto \exp(-E_f/k_B T)$). Hand et al. proposed that beyond a certain threshold temperature $T_t$ the absorption will be abruptly high, which supports PFF. Later coming studies extended this theory. Yakovlenko et al. deduced the speed of PFF in a 2D thermal absorption wave theory but did not use such an absorption relation as Eq. 1 at first [31, 32]; an equation similar to Eq. 1 was used later in including plasma into the physics of PFF [33]. Shuto et al. [34, 35] used Eq. 1 to simulate evolution of temperature field during PFF, using a 2D heat transfer model with a linear heat source determined by the absorption. Both $E_f$ and $\alpha_0$ had to be chosen for letting the simulated evolution speed of temperature field match a speed of PFF in experiment. There are studies that used different equations [8, 36-39] to simulate the absorption, for example, a step-function form [36, 37] as $\alpha(T) \propto T$ for some ranges $T_1 < T < T_P$, or simply $\alpha(T) \propto T^4$ for all temperatures [8, 38]. The studies all obtained results in agreement with the experimental speeds of PFF.

Presuming the intrinsic absorption, the above studies can well describe some features of PFF. Based on that, it is possible to extend the defect-induced absorption theory to IFF as a natural extension of the physical process. But it is noteworthy that the discrepancies among the details of the above successful theories are still fundamental. Different combinations of the details respectively obtained results plausibly agreeing with the few and indirect experimental characteristics (such as the speed of PFF), while only one combination can be physically true. There is a lack of more direct method to uniquely determine many of the details, including the true mathematical form of the absorption, and the true values of key parameters such as $E_f$ and $\alpha_0$, $\alpha_P$ and $T_1$, etc. Most importantly, there is little quantitative information about IFF that can be deduced from the above theories. This cannot meet the need for preventing spontaneous IFF. In summary, there are two pieces missing in the current puzzle of IFF: more direct experimental characteristics, and a method in which more details can be uniquely determined.

## 2.2 Experimental results of the critical conditions for IFF

If IFF is an intrinsic property of silica fibers, it will have some critical conditions. In our previous study [29], the critical conditions for IFF were measured for the first time; the results related to later analyses of this paper are briefly reviewed here. A critical temperature $T_c$ for IFF is presumed, which let the temperature of the fiber material enters an unstoppably rising pattern. As laser light supports the fiber fuse process, $T_c$ is likely to vary with a critical optical power $P_c$. At each time of measurement, a fiber was uniformly heated till IFF happened; the setup is schemed in Fig. 2. The recorded temperatures at the moments of IFF were used as $T_c$. The recorded output optical powers at the beginning of heating ($P_{in}$) and at the moments of IFFs ($P_{out}$) were used to calculate the critical optical powers $P_c = \sqrt{P_{in}P_{out}}$. The experiment was costly. Each time averagely, a ~2-m fiber had to be destroyed, including failed times in which no fiber fuses were initiated but the fiber was still functionally damaged. For relative cost-effectiveness, we only tested single-mode silica fibers, including 4 kinds of 10/130-μm YDFs, 1 kind of 10/130-μm passive (with no active ions doped) fiber and 1 kind of G652.D passive fiber.

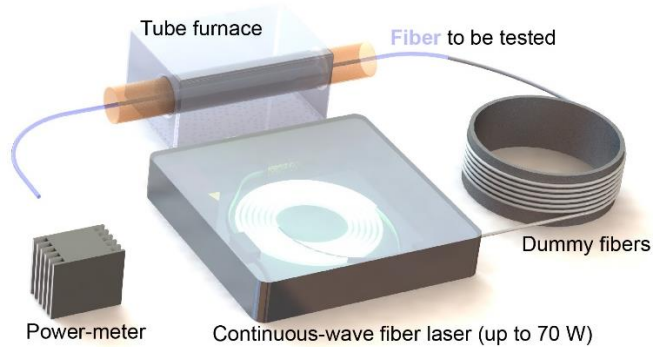

Fig. 2. Experimental setup for measuring the critical conditions for IFF.

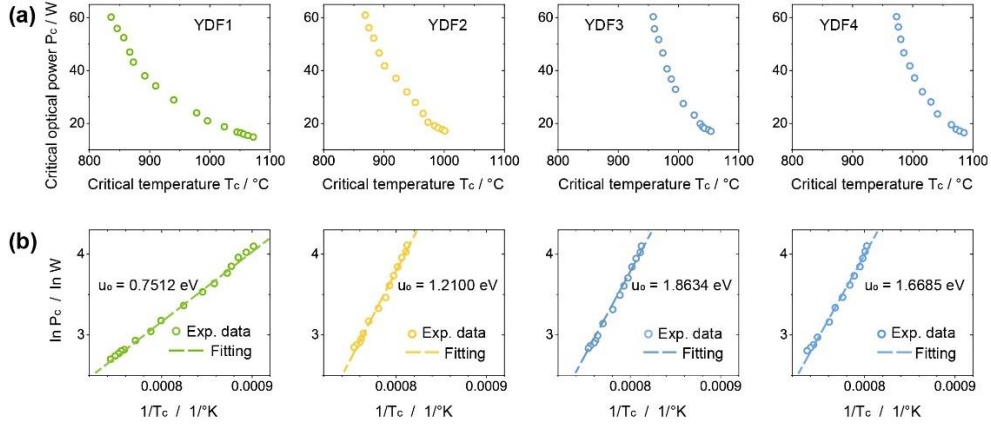

Fig. 3. Experimental data of the critical conditions for IFF. (a) Critical temperatures $T_c$ and critical optical powers $P_c$ for IFF. (b) Showing the linear correlation between $1/T_c$ and $\ln P_c$.

As a result, none of the tested passive fibers had any IFF even under 1,200°C. In fact, this is in agreement with previous report that passive fibers did not have fiber fuses under ~1,000°C when heated in the middle [40]. In contrast, all the 4 kinds of tested YDFs had IFFs in some ranges of applied conditions. Data of the measured critical conditions $T_c$ and $P_c$ were uploaded in the supplementary information of the previous paper [29] and are here illustrated in Fig. 3(a), where each data point represents an experimentally triggered IFF. There is a correlation between the critical temperatures and critical optical powers for IFFs in all the fibers. By taking logarithm for $P_c$ and inversion for $T_c$, a clear linear correlation $\ln P_c \propto 1/T_c$ is revealed, as shown in Fig. 3(b). (It is noted that the values of $u_0$ shown in the previous paper [29] were erroneous due to calculation errors; the correct values are shown here, as can be re-calculated using the original experimental data in the previous paper.) Moreover, if we write the linear correlation in the following form

$$\ln P_C = \frac{u_0}{k_B} \times \frac{1}{T_C} + \ln \gamma, \qquad (2)$$

where $u_0$ is a parameter with the dimension of energy, $\gamma$ is another parameter with the same dimension as $P_C$, the correlation in Fig. 3(a) can be written as

$$P_C = \gamma \cdot \exp\left(\frac{u_0}{k_B T_C}\right). \qquad (3)$$

This equation looks familiar with Eq. 1, but the negative sign in the exponent term in Eq. 1 is not present here. In our previous study, we flipped $P_C$ and $\gamma$ and had $\gamma = P_C \exp(-u_0/k_B T_C)$, which had the same form as Eq. 1, but the physical meaning of the equation was not clear.

Here, we propose a new explanation. Suppose there be a kind of Frenkel defect in the solid-state fiber materials that causes intrinsic absorption and leads to IFF, then the concentration of the defects $n_F$ at temperature $T$ is

$$n_F(T) = n_0 \exp\left(-\frac{u_F}{k_B T}\right), \qquad (4)$$

where $u_F$ is the formation energy of the defect from the original (equilibrium) crystal cube, $n_0$ is the upper limit of the concentration when $T \to \infty$. If we presume that the absorption α is linearly dependent on $n_F$, it can be written as

$$\alpha(T) = \alpha_F n_0 \exp\left(-\frac{u_F}{k_B T}\right) = a_0 \exp\left(-\frac{u_F}{k_B T}\right) \propto n_F(T), \qquad (5)$$

where $\alpha_F$ can be understood as the probability (or efficiency) of that a defect absorb optical power, $a_0 = \alpha_F n_0$ is a simplified coefficient. So, for IFF, if $T_{\text{core}}$ is the temperature of the fiber core, the absorbed power $P_{\text{ab},C}$ will be the product of $\alpha(T_{\text{core}})$ and $P_C$, as

$$P_{\text{ab},C} = \alpha_0 \exp\left(-\frac{u_F}{k_B T_{\text{core}}}\right) \cdot \gamma \exp\left(\frac{u_0}{k_B T_C}\right) = \alpha_0 \gamma \exp\left(\frac{1}{k_B}\left(\frac{u_0}{T_C} - \frac{u_F}{T_{\text{core}}}\right)\right). \quad (6)$$

Here, a luring hypothesis is that IFF (in a certain fiber) requires an almost invariant critical destructive power $P_{\text{ab},C}$ from absorption, which is written as $P_{\text{ab},C} \approx \alpha_0 \gamma$. This assumption is plausible, as damage occurs on the chemical level so that the required energy for the initial decomposition of material (the IFF) should be constant. This constant requires

$$\frac{u_0}{T_C} - \frac{u_F}{T_{\text{core}}} \approx 0. \quad (7)$$

In our previous paper, $T_C \approx T_{\text{core}}$ was implicitly adopted as a good approximation, which means the measured $T_C$ should be close enough to $T_{\text{core}}$. If $T_C \approx T_{\text{core}}$ is true, Eq. 7 further demands $u_F \approx u_0$, which means the experimentally revealed parameter $u_0$ is directly the formation energy of chemical defects responsible for IFFs. In our previous paper, we implicitly acknowledged that $u_F = u_0$, but it could not be proved at that time.

In the next chapter, we use numerical analysis to verify if $u_F = u_0$ is true by verifying if simulation results based on $u_F = u_0$ can agree with the direct experimental data (the critical conditions for IFF). As shown later, it turns out that $u_F = u_0$ is not accurate: $u_0$ is not the formation energy of chemical defects, but is always slightly smaller than that. The analysis not only uniquely determines $u_F$ for each YDF, but also prove that a defect-induced absorption is a major mechanism for IFF. The determined $u_F$ is critical information for further studying the physical mechanisms of IFF at material and chemical level in future.

## 3. Verifying the defect-induced absorption mechanism of IFF

### 3.1 3D solid-state heat transfer model with absorption-induced heat source

The key parameter of the theoretical model for IFF will be the temperature. As aforementioned, a simulated temperature unstoppably rising means that the applied initial conditions surpassed the critical conditions for IFF; otherwise, a simulated temperature stabilizing at lower than the softening point of silica fibers means that the initial conditions were lower than the critical conditions for IFF. In this way, simulation can find the critical conditions for IFF by repeatedly trying to narrow the range between higher and lower results to a tolerable level. In this model, only $u_F$ and $\alpha_0$ await selection as objected optimization parameters. For each YDF, there are plenty data of the critical conditions, so only when one invariant group of $u_F$ and $\alpha_0$ achieve agreement with all the experimental data, will it mean that the selected values of $u_F$ and $\alpha_0$ are the true values (that their existence is physically self-consistent). Moreover, if that can be achieved, it will also prove that the model here reflects the physical truth of the IFF, that IFF is caused by defect-induced absorption under solid-state heat transfer.

Before material phase transition of fiber in an IFF, the material should stay solid state. So, it should be available to simulate the real-time temperature evolution before IFF with a 3D heat transfer model. Consider a general case that the model is symmetrical around the axis of fiber; all simulation variables are functions of radial coordinate $r$ and axial coordinate $z$. The geometry of the model is set according to experiment, 3 coaxial cylindrical layers of material represent respectively fiber core, fiber inner cladding, and air, as shown in Fig. 4. Their diameters are chosen to be 10, 130 and 330 μm, respectively. Considering the small heating region and transient process of IFF in the experiment, the simulation model uses solid-state heat transfer, neglecting the fluid effects of air. The heat source is the defect-induced absorption of laser light transmitted in fiber to adopt the above absorption mechanism. In principle, a full numerical treatment needs coupling between electromagnetic equations with the heat transfer equations. However, for an accurate-enough approximation, we suppose the intensity of optical field propagates towards $z+$ and always remains a Gaussian profile in all the cross-sections of fiber along $z$. Optical power $P_{\text{in}}$ (W) is input at cross-section $z = 0$. So, the area density of optical power $P(r, z = 0)$ (W/m²) in the 3D space is

$$P(r,0) = \frac{P_0}{\pi r_0^2} \exp\left(-\frac{2r^2}{r_0^2}\right), \tag{7}$$

where $r_0$ is the radius of the Gaussian optical field. In each cross section $z$, $P(r,z)$ equals $P(r, z-d\xi)$ multiplying $1 - \alpha(r,z)d\xi$, where $d\xi$ is an infinitesimal distance along $z$, and $\alpha(r,z)$ (m$^{-1}$) is the defect-induced absorption defined by Eq. 5. So, $P(r,z)$ is the limit of a continued multiplication

$$\begin{aligned} P(r,z) &= P(r,0) \lim_{d\xi \to 0} \prod_{j=1}^{\frac{z}{d\xi}} (1 - \alpha(r, j \cdot d\xi)d\xi) \\ &= P(r,0)\exp\left(-\int_0^z \alpha(r,\xi)d\xi\right). \end{aligned} \tag{8}$$

The volume density of heat source $\dot{Q}_{ab}(r,z)$ (W/m$^3$) is $P(r,z)$ multiplied by $\alpha(r,z)$,

$$\dot{Q}_{ab}(r,z) = \alpha(T)P(r,z) = P(r,z)\alpha_0 \exp\left(-\frac{u_F}{k_B T}\right), \tag{9}$$

where $u_F = u_0$ is used at first, according to the assumption in the last chapter.

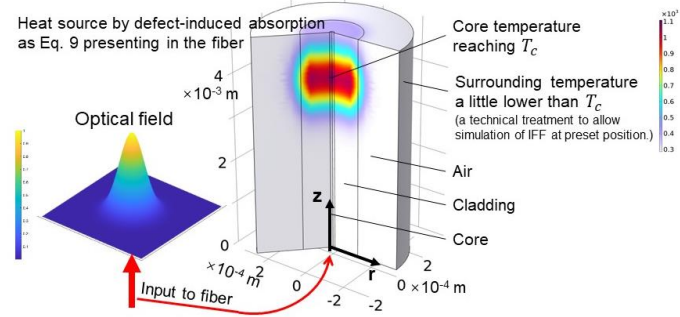

Fig. 4. Model of simulation using the 3D solid-state heat transfer with defect-induced-absorption heat source.

To simulate IFF, an initial high-temperature field is applied, then temperature evolution with time is solved numerically by using Eq. 9 to update the heat source in every time step. If temperature rises unstoppably and exhibits an accelerated increasing trend (towards infinity), it means IFF happens under such high-temperature initial conditions. Otherwise, if temperature stopped rising at a finite value below the melting point of silica material, it means no IFF happens. Whether the physical implications (including $u_F = u_0$) sustain depends on whether IFFs happen (and just enough to happen) in simulation by using the experimentally measured critical conditions for IFF. According to our experimental setup, the initial high-temperature field for simulation can be

$$T(z)_{t=0} = T_c - \Delta T + \Delta T \exp\left(-\frac{(z-z_0)^2}{(\Delta z)^2}\right), \tag{10}$$

where $T_c$ is the experimentally measured critical temperature for IFF. $\Delta T$ is a perturbation of temperature, $z_0$ is location of the maximum of the initial temperature field and $\Delta z$ is the width of the perturbed temperature field. The reason for introducing such perturbations $\Delta T$, $z_0$ and $\Delta z$ is that we want to let IFFs in simulation happen around $z_0$, which mimics the experimental cases where small temperature gradients existed in the tube furnace in Fig. 2 and made IFFs always happen at the middle point of the heated fibers. If no perturbation is present and instead a constant initial temperature is used, then IFFs will always happen at $z = 0$ where $P(r,z)$ has a maximum, according to Eq. 8. About the optical power $P(r,0)_t$, Eq. 7 uses

$$P_0 \equiv P_c, \tag{11}$$

where $P_c$ is the experimentally measured critical optical power for IFF, as it is the optical power transmitted through the local area of IFF.

Till now, the only unknown parameter is the absorption coefficient $\alpha_0$. If the model is physically true, the 3 parameters $\alpha_0$, $T_c$ and $P_c$ are correlated. When the true value of $\alpha_0$ is selected, $T_c$ and $P_c$ will naturally be solutions of simulation. So, the strategy to obtain $\alpha_0$ is as follows. First, adopt a random value for $\alpha_0$. Then, use a pair of experimentally measured $T_c$ and $P_c$ (a data point in Fig. 3a) to initiate the simulation, see if IFF happens in simulation under such conditions. If IFF happens, it means that $\alpha_0$ is enough and, importantly, probably too high for IFF, the true $\alpha_0$ should be smaller; then, use a smaller $\alpha_0$ to repeat again. If IFF does not happen, it means $\alpha_0$ is too low for IFF, the true $\alpha_0$ should be bigger; then, use a bigger $\alpha_0$ to repeat again. Keep repeating till the precision of the selected value of $\alpha_0$ reaches a certain level.

After an $\alpha_0$ is determined for a pair of experimental $T_c$ and $P_c$, the physical implication will be that the defect-induced absorption under $\alpha_0$ accurately causes IFF under the conditions $T_c$ and $P_c$. Then, by each time substituting a different $P_c$ into the simulation with $\alpha_0$, we can determine a new $T_c$. If all our above assumptions were correct (the defect-induced absorption was true, and $u_F = u_0$ was also true), the simulated $P_c$ and $T_c$ should all agree with all the experimentally measured data.

### 3.2 Determining the true values of $u_F$

As explained before, we are using $u_F = u_0$ for Eq. 9 at first. We simulate $\alpha_0$, $P_c$ and $T_c$ for comparison with the experimental data. For each kind of YDF, we choose first the highest data point (of the highest critical optical power $P_c$) in Fig.3a and use it ($P_c$ and $T_c$) to determine $\alpha_0$, then we use the determined $\alpha_0$ and all the $P_c$ to simulate $T_c$. Then, according to the steps above, we choose another data point and repeat the simulation likewise. The results are shown in Fig. 5a for comparison. As can be seen, the simulated $P_c$ and $T_c$ always clearly deviate from the experimental data; this effect is most conspicuous in YDF 1. Using the initially chosen data point of $P_c$ and $T_c$ as a baseline, at larger $P_c$, the resulted $T_c$ will be higher than the experimental data; while using a smaller $P_c$, $T_c$ will be smaller. The deviation implies that some part of the simulation may be incorrect. In fact, by using $u_F = u_0$, the deviation persists no matter what values of $\alpha_0$ is applied.

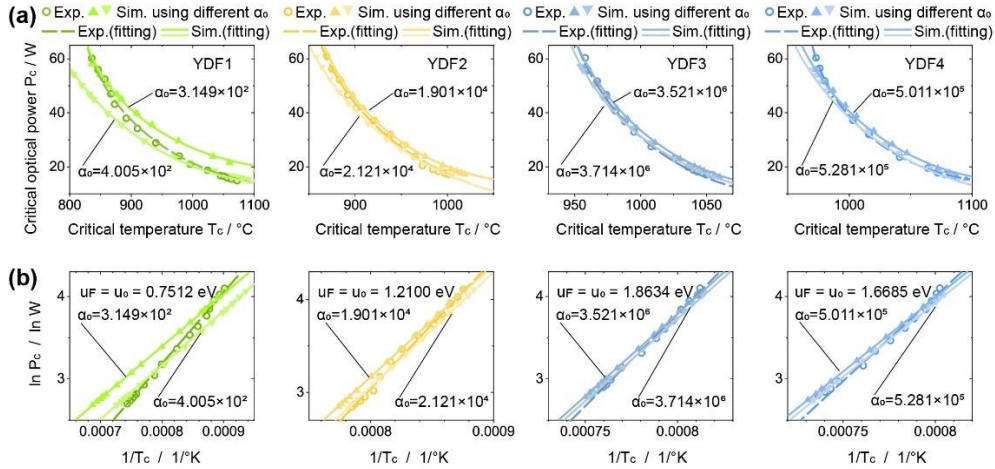

Fig. 5. Simulation results of the critical conditions, using $u_F = u_0$, in comparison with the experimental results in Fig. 3; note that the horizontal axis here no longer remain of the same scales as in Fig. 3. (a) Simulation results using $u_F = u_0$ only obviously deviate from the experimental results. (b) Simulation results in coordination transformation as in Fig. 3(b).

In spite of the deviation, however, it is clear that the simulation results of $P_c$ and $T_c$ have the same characteristics (pattern of variation) as the experimental data. In fact, using the same transformation of coordinates as Fig. 3b, it can be seen in Fig. 5b that the simulated $P_c$ and $T_c$ still have the distinct correlation $\ln P_c \propto 1/T_c$. It's just that the slopes of the curves of simulation results are somehow smaller than that of the experimental results. The existence of the same correlation pattern suggests that the main part of the model may be fine and it is likely that values of some important parameters have had systematic errors.

In fact, it is soon realized that the mismatch may come from a deviated value of $u_F$. As can be understood, if using $u_F = u_{0,\exp}$ can cause $u_{0,\text{sim}} < u_{0,\exp}$, for compensating such deviation, perhaps using $u_F > u_{0,\exp}$ will work (causing $u_{0,\text{sim}} = u_{0,\exp}$). The principle of this process can be shown in Fig. 6. By increasing $u_F$, the slopes of all the curves of simulation results in the lower row of Fig. 6 will increase. It is foreseeable that, when $u_F$ is sufficiently large, the curves of simulation results will have the same slopes as that of the experimental results. Then, there must be a specific $\alpha_0$ that perfectly align simulation results to the experimental data. In this way, $u_F$ and $\alpha_0$ will allow physically self-consistent results for all the simulation inputs, then it means they are finally determined.

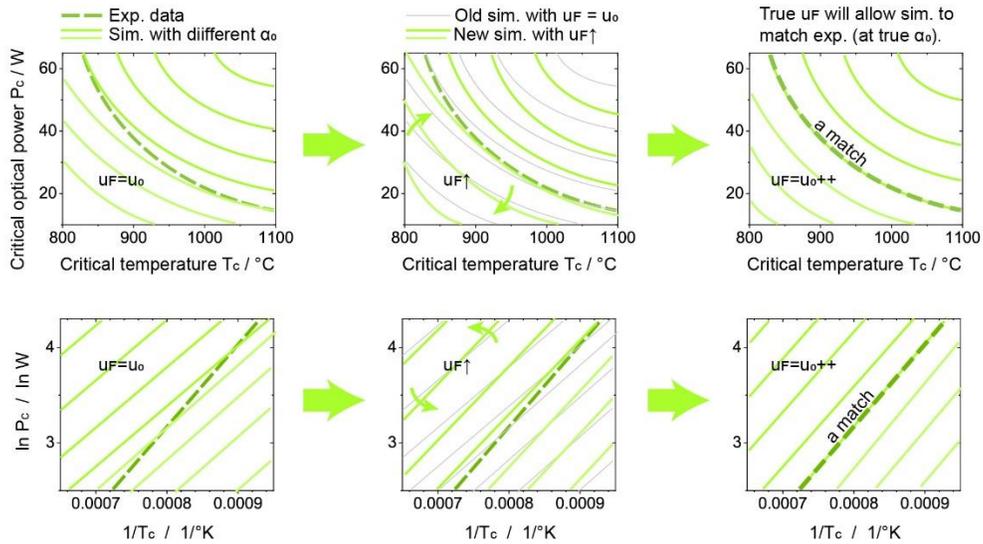

Fig. 6. Principles of determining the true value of $u_F$ (showing YDF1 as an example); all the curves of simulation results in this figure are schematic with no specific values. Subplots on the left show that initially by using $u_F = u_0$ the simulation results will always deviate from the experimental results. Subplots in the middle show that by increasing $u_F$ the curves of simulation results will obtain larger slopes; the simulation results will attain a closer pattern of variation to that of the experimental results. Subplots on the right show that when true value of $u_F$ is met, it will allow a match between the experimental results and some simulation results at true $\alpha_0$. Till this, the true value of $u_F$ (as well as $\alpha_0$) is determined.

Based on the above principles, we adjust the values of $u_F$ to repeat the whole simulation (from $\alpha_0$ to $T_c$), till we find simulation results of $P_c$ and $T_c$ in good agreement with all the experimental data. In fact, this can be done by taking a shortcut that the value of $u_F$ is close to the experimental $u_{0,\exp}$ adding its difference with the previously simulated $u_{0,\text{sim}}$, written as

$$u_F \approx u_{0,\exp} + (u_{0,\exp} - u_{0,\text{sim}}) = 2u_{0,\exp} - u_{0,\text{sim}}. \tag{12}$$

After several rounds of tries, the self-consistent true values of $u_F$ can be found. The simulated results of $u_F$, $\alpha_0$, $P_c$ and $T_c$ for the 4 kinds of YDFs are shown in Fig. 7, in comparison with the experimental data. It is obvious that this time the simulation results are in good agreement with the experimental results. The fact that $u_F > u_0$ is explained later. The agreement here suggests that the defect-induced absorption mechanism can well describe IFF in predicting its critical conditions. In this way, it can be recognized that the defect-induced absorption is proved.

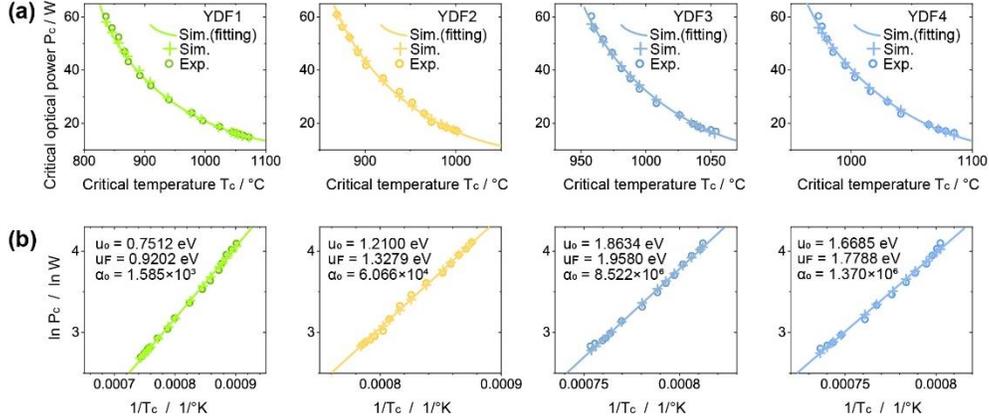

Fig. 7. Simulations results that determine the true values of $u_F$ and $\alpha_0$ by best matching the experimental results of the critical conditions for IFF. (a) Simulation results comparing with data in Fig. 3(a). (b) That comparing with Fig. 3(b).

## 4. Analytical solution of the steady-state temperature of gain fiber

### 4.1 General analytical solution

The above results suggest that the risk of fiber fuse is related to the operating temperatures and optical powers of the fibers. In Chap. 3, we used numerical simulation to solve the temperature of fiber under the measurement setup, which was a structurally specific and simple case with basically no involvement of laser dynamics in YDFs. However, for more complex and laser-dynamics-involved cases, it will be complicated to code the numerical simulation to couple and customize the multiple physical processes. At this point, an analytical solution will be useful for evaluating the temperature of gain fiber and thus the risk of fiber fuse. However, few analytical solutions from previous studies elsewhere can satisfy the need here. There are two major reasons. First, most previous solutions treat heat source as either a line along the symmetrical axis or a uniform distribution in the whole space of the model, which lacks precision for further analyzing the steady-state operating status. Second, few previous solutions treat heat source varying along the whole fiber while coupled with laser gain process.

To obtain a more general analytical solution of the steady-state operating temperatures of gain fibers, we start from a 3D solid-state heat transfer model with the heat transfer equation

$$\rho c \frac{\partial T}{\partial t} = \nabla \cdot (k \nabla T) + \dot{Q}_l, \tag{13}$$

where $\rho$ is density of mass (kg/m³), $c$ (constant volume) specific heat, $k$ heat conductivity (W/(m·K)) and $\dot{Q}_l$ the density of power of heat source. Similar to the last chapter, we use cylindrical coordinates $(r, \varphi, z)$ to rewrite the equation as

$$\rho c \frac{\partial T}{\partial t} = \frac{1}{r}\frac{\partial}{\partial r}\left(kr\frac{\partial T}{\partial r}\right) + \frac{1}{r^2}\frac{\partial}{\partial \varphi}\left(k\frac{\partial T}{\partial \varphi}\right) + \frac{\partial}{\partial z}\left(k\frac{\partial T}{\partial z}\right) + \dot{Q}_l. \tag{14}$$

For model of multiple materials, $\rho$, $c$, $k$, $\dot{Q}_l$ and $T$ are variables of the coordinates. But considering rotational symmetry, the heat transfer around the axis $z$ is absent, so $\partial/\partial \varphi = 0$. In continuous-wave fiber lasers, such as high-power ytterbium-doped fiber lasers (YDFLs), the operating optical power and the underlying laser mechanism in ideal situation are steady-state. So, the heat source in YDFLs generated by quantum deficit, as well as the operating temperature, is also steady-state. The relatively slowly varying heat source along $z$ allows us to neglect axial heat transfer, so $\partial/\partial z = 0$. Note that coordinate $z$ still exists. Moreover, as what we are interested in is the steady-state temperature $T(t \to \infty)$ where $\partial/\partial t = 0$, we have

$$\frac{1}{r}\frac{d}{dr}\left(kr\frac{dT}{dr}\right) = -\dot{Q}_l, \tag{15}$$

where $\partial r$ is replaced by $dr$ for simplicity. For our model of multiple layers of materials along $r$ (so $k$ and $\dot{Q}_l$ are not constant with all the coordinates), this equation must be solved by integration. The first time of integration gives

$$\int d\left(kr\frac{dT}{dr}\right) = -\int \dot{Q}_l r dr = -\left(\int d\left(\dot{Q}_l \frac{r^2}{2}\right) - \int \frac{r^2}{2} d\dot{Q}_l\right). \tag{16}$$

The change from $dr$ to $d\dot{Q}_l$, as well as that from $dr$ to $dk$ (Eq. 18), is a deliberate move. It is because that the model aims to let $\dot{Q}_l$ and $k$ be layered in space so $d\dot{Q}_l$ and $dk$ only exist at the surfaces between the layers; this merits simplification from continuous integration to discrete addition, which leads us to a general analytical solution. This is a key step to obtain a general analytical solution without involving any model-specific subsection integration.

Now, before continuing the integration, Eq. 16 requires specifying the field of integration. For simplifying the following calculation, we can choose from 0 to $r$ as the field, because at 0 there must be $kr(dT/dr) = 0$ and $\dot{Q}_l r^2/2 = 0$. So, Eq. 16 simplifies to

$$kr\frac{dT}{dr} = -\left(\dot{Q}_l \frac{r^2}{2} - \int_0^r \frac{r^2}{2} d\dot{Q}_l\right). \tag{17}$$

Note that the latter term on the right-hand side of Eq. 17 is now a constant in fact; there is no undetermined constant generated because it is a definite integration from Eq. 16 to Eq. 17, where both sides of the equation are integrated. Again, an integration of Eq. 18 gives

$$\int dT = -\frac{1}{2} \int \frac{\dot{Q}_l}{k} r dr + \left(\int_0^r \frac{r^2}{2} d\dot{Q}_l\right) \int \frac{dr}{kr}$$

$$= -\frac{1}{2}\left(\int d\left(\frac{\dot{Q}_l}{k}\frac{r^2}{2}\right) - \int \frac{r^2}{2} d\frac{\dot{Q}_l}{k}\right) + \left(\int_0^r \frac{r^2}{2} d\dot{Q}_l\right) \int \frac{d\ln r}{k}$$

$$= -\frac{1}{4}\left(\int d\left(\frac{\dot{Q}_l r^2}{k}\right) - \int r^2 d\frac{\dot{Q}_l}{k}\right) + \left(\int_0^r \frac{r^2}{2} d\dot{Q}_l\right) \int \left(d\frac{\ln r}{k} - \ln r\, d\frac{1}{k}\right). \tag{18}$$

The change from $dr$ to $dk$ ($d(1/k)$) is deliberate, as explained above. Likewise, by integrating from 0 to $r$, Eq. 18 gives

$$T(r) = T(0) - \frac{1}{4}\left(\frac{\dot{Q}_l r^2}{k} - \int_0^r r^2 d\frac{\dot{Q}_l}{k}\right) + \left(\int_0^r \frac{r^2}{2} d\dot{Q}_l\right) \int_0^r \left(d\frac{\ln r}{k} - \ln r\, d\frac{1}{k}\right). \tag{19}$$

This is the general analytical solution of the operating temperature of fiber under 3D solid-state heat transfer. It is applicable without requiring specific information about the numbers and the sizes of the layers. Thus, it can be used to deduce final analytical solutions for various kinds of model structures, as long as the number of discrete points where $d\dot{Q}_l$ or $d(1/k)$ equals non-zero is finite.

### 4.2 Setup of heat source from quantum deficit in continuous-wave YDFL

After obtaining the analytical solution, what needs to be considered next is the heat source $\dot{Q}_l(r,z)$. In safely operating high-power continuous-wave YDFLs, $\dot{Q}_l(r,z)$ and the underlying quantum deficit are expected to be steady-state (time-invariant). As the extraction efficiency is high, it can be seen that an annihilated pump photon corresponds to a generated signal photon in the fiber. If we can solve the optical power distribution $P_{+/-}(z)$ (W) of the co-propagating (+) and counter-propagating (-) light along fiber, the distribution of heat source $\dot{Q}_2(z)$ (W/m) should take the following form

$$\dot{Q}_2(z) = c_{\text{qd}}\left(\frac{dP_+(z)}{dz} + \frac{dP_-(z)}{dz}\right), \tag{20}$$

where $c_{\text{qd}}$ is a coefficient of quantum deficit, which is the proportion of the lost energy from a pump photon $\lambda_p$ to a signal photon $\lambda_s$ in the energy of the signal photon, written as

$$c_{\text{qd}} = \left(\frac{1}{\lambda_p} - \frac{1}{\lambda_s}\right) \Big/ \frac{1}{\lambda_s} = \frac{\lambda_s}{\lambda_p} - 1, \tag{21}$$

where $\lambda_{p/s}$ is the wavelength of the pump/signal photon. In fact, for a multiple-wavelength broadband simulation model that solves a spectral power distribution $P^{\lambda}_{+/-}(z)$, Eq. 21 is written as

$$\dot{Q}_2(r,z) = \sum_{\lambda_s}\left(\{c_{qd}^{\lambda_s}\}\cdot\left(\left\{\frac{dP_+^{\lambda_s}}{dz}\right\}+\left\{\frac{dP_-^{\lambda_s}}{dz}\right\}\right)\right), \quad (22)$$

where $\{c_{qd}^{\lambda_s}\}$ is a column vector with the same arrangement of element as $\{P^{\lambda_s}_{+/-}\}$, dot product '·' is multiplication by element.

Now, to change $\dot{Q}_2$ (W/m) into $\dot{Q}_l$ (W/m³), the area density $P_{+/-}(r,z)$ needs to be known. In fact, for our experimentally used YDFs, it is possible to assume that

$$P_{+/-}(r,z) = \begin{cases} \dfrac{P_{+/-}(z)}{\pi r_0^2}, & 0 \le r \le r_1 \\ 0, & r > r_1 \end{cases}, \quad (23)$$

where $r_0$ is mode radius and $r_1$ the radius of the doped core of YDF. Then,

$$\dot{Q}_l(r,z) = \begin{cases} c_{qd}\left(\dfrac{dP_+(z)}{dz}+\dfrac{dP_-(z)}{dz}\right)\dfrac{1}{\pi r_0^2}, & 0 \le r \le r_1 \\ 0, & r > r_1 \end{cases} = \begin{cases} \dot{Q}_1, & 0 \le r \le r_1 \\ 0, & r > r_1 \end{cases}. \quad (24)$$

After obtaining the heat source $\dot{Q}_1$, we can further setup the model. For typical cases, we assume that there are 5 layers (core, inner cladding, coating, heat conducting glue and heat sink) with radii $0 < r_1 < r_2 < r_3 < r_4 \le r_5 < \infty$, the heat conductivities $k_1 \sim k_5$ of the respective layers are invariant with temperature. At this point, one may hold on because in real situations, such as high-power fiber lasers, the last two layers ($k_4$ and $k_5$) around the fiber is usually not circularly symmetrical, for example as shown in Fig. 8 (left). However, we propose that the non-symmetrical case can be easily approximated by a circularly symmetrical case shown in Fig. 8 (right) with fair accuracy, by using $k_4/2$ and $k_5/2$ to replace the value of $k_4$ and $k_5$ respectively in calculation. By using this approximation

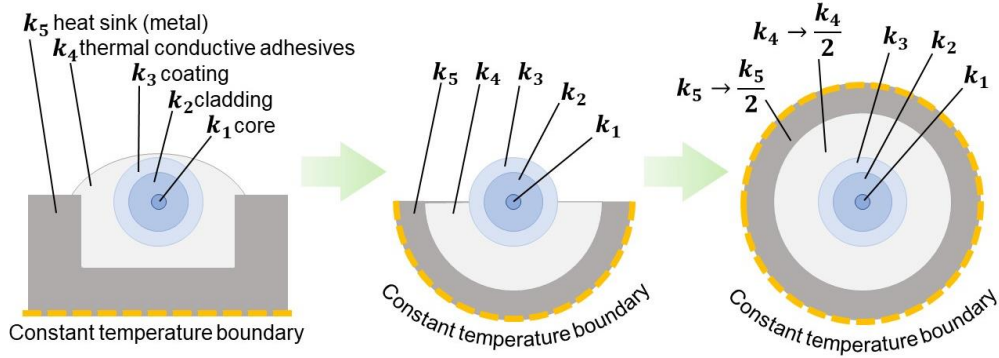

Fig. 8. Typical structure of the transverse structure around a fiber embedded in cooling setup (left) and its approximation for a cylindrically symmetrical model (right).

Let the boundary temperature condition be
$$T(r_5) = T_{\text{boundary}} = T_5. \quad (25)$$

By substituting this into Eq. 19, noting that $\dot{Q}_1$ only exists in $0 \le r \le r_1$, it gives

$$T(0) = T_5 + \frac{\dot{Q}_1 r_1^2}{4k_1} + \frac{\dot{Q}_1 r_1^2}{2}\left(\frac{\ln\frac{r_5}{r_4}}{k_5} + \frac{\ln\frac{r_4}{r_3}}{k_4} + \frac{\ln\frac{r_3}{r_2}}{k_3} + \frac{\ln\frac{r_2}{r_1}}{k_2}\right). \quad (26)$$

This is the analytical solution of the maximum temperature $T(0)$ in a cross section of YDF. For other positions $r$, by substituting Eq. 26 back into Eq. 19, it gives

$$T(r) = \begin{cases} T_5 + \dfrac{\dot{Q}_1}{4k_1}(r_1^2 - r^2) + \dfrac{\dot{Q}_1 r_1^2}{2}\left(\dfrac{\ln\frac{r_5}{r_4}}{k_5} + \dfrac{\ln\frac{r_4}{r_3}}{k_4} + \dfrac{\ln\frac{r_3}{r_2}}{k_3} + \dfrac{\ln\frac{r_2}{r_1}}{k_2}\right), & 0 \leq r < r_1 \\[6pt] T_5 + \dfrac{\dot{Q}_1 r_1^2}{2}\left(\dfrac{\ln\frac{r_5}{r_4}}{k_5} + \dfrac{\ln\frac{r_4}{r_3}}{k_4} + \dfrac{\ln\frac{r_3}{r_2}}{k_3} + \dfrac{\ln\frac{r_2}{r}}{k_2}\right), & r_1 \leq r < r_2 \\[6pt] T_5 + \dfrac{\dot{Q}_1 r_1^2}{2}\left(\dfrac{\ln\frac{r_5}{r_4}}{k_5} + \dfrac{\ln\frac{r_4}{r_3}}{k_4} + \dfrac{\ln\frac{r_3}{r}}{k_3}\right), & r_2 \leq r < r_3 \\[6pt] T_5 + \dfrac{\dot{Q}_1 r_1^2}{2}\left(\dfrac{\ln\frac{r_5}{r_4}}{k_5} + \dfrac{\ln\frac{r_4}{r}}{k_4}\right), & r_3 \leq r < r_4 \\[6pt] T_5 + \dfrac{\dot{Q}_1 r_1^2}{2}\left(\dfrac{\ln\frac{r_5}{r}}{k_5}\right), & r_4 \leq r \leq r_5 \end{cases} \quad . \quad (27)$$

This is the analytical solution of steady-state operating temperature of a fiber in the 5-layered model; note that coordinate $z$ is implicit. This analytical solution is based on the 2 above-mentioned approximations: (1) uniform distribution of only-in-core heat source $\dot{Q}_1$, and (2) layered materials with heat conductivity of each material being invariant with temperature. If any of these two fails, $d\dot{Q}_l$ or $dk$ will no longer be discretized in space so that the integrations in Eq. 25 cannot be analytically simplified, and the solution cannot be obtained. For most cases, including the later discussed high-power YDFLs, these 2 approximations are good enough to provide an overlook of the potential limits of performance.

### 4.3 Reviewing the result that $u_F > u_0$

The above analytical solution actually tells why $u_0$ revealed in experiment is a little smaller than the suitable $u_F$ that produce simulation results precisely matching the experimental results. For seeing that, use Eq. 33 (for the absent layers, use $r_3 = r_4 = r_5$ for example) to give

$$T_{\text{core}} = T(0) = T_{\text{boundary}} + c_0 \dot{Q}_1, \tag{28}$$

where $c_0$ represents the functions of all the structural layers, written as

$$c_0 = \frac{r_1^2}{2}\left(\frac{\ln\frac{r_5}{r_4}}{k_5} + \frac{\ln\frac{r_4}{r_3}}{k_4} + \frac{\ln\frac{r_3}{r_2}}{k_3} + \frac{\ln\frac{r_2}{r_1}}{k_2} + \frac{1}{2k_1}\right). \tag{29}$$

Here, what can be measured in experiment is $T_{\text{boundary}}$, while what actually causes IFF is $T_{\text{core}}$. In the experimental results in Fig. 3, $u_0$ was fitted using $T_{\text{boundary}}$ as the critical temperature $T_C$ for IFF. Thus, we can use Eq. 7 and Eq. 34 to obtain (note that $\dot{Q}_1$ can slightly vary with $T_C$ in a complex relation in the range of experimental parameters)

$$u_F = \frac{T_{\text{core}}}{T_C} u_0 = (1 + \frac{c_0 \dot{Q}_1}{T_C}) u_0 > u_0. \tag{30}$$

This equation means that $u_F > u_0$ holds all the time, which explains the principles in Fig. 6 and the results in Fig. 7, as well as Eq. 12. In our previous study, we neglected the difference between $T_c$ and $T_{\text{core}}$. Then, we were stuck in the hypothesis that $u_0 = u_F$ and could not obtain further results. Equation 30 suggests that the experimentally revealed $u_0$ is not directly the formation energy of lattice defects. It is, however, a comprehensive effect from the formation of defects and the heat dissipation structure of the fiber. By iterating the simulation, we can obtain the real $u_F$ with self-consistent results; using the same simulation parameters, simulated critical conditions for IFF can well match the experimental results. The agreement proves that the defect-induced absorption is a major mechanism of IFF.

## 5. Steady-state operating status of YDFLs using the 10/130-μm YDF

The general analytical solutions of the steady-state operating temperatures of gain fibers can be used to analyze the operating status of fiber lasers, namely the operating temperatures and the operating optical powers of all positions along the fibers. This analysis is useful, as the results can be compared with the critical conditions for IFF and it can tell whether an operating status is safe. Here we present an example of such analysis. We consider a continuous-wave YDF amplifier in a co-pumping configuration. Amplifier is a key part of high-power YDFLs, most of which use master-oscillator-power-amplifier (MOPA) structure; it usually operates with high thermal loads and high probability of spontaneous fiber fuse damage in reality. For later comparisons with the experimental data of the critical conditions for IFF, we consider 10/130-μm YDFs.

In state-of-the-art high-power YDFLs, one of the following two pumping wavelengths is often used: conventional LD pumping at ~976 nm, and tandem pumping (pumping by fiber lasers) at ~1018 nm [2, 16]. The 1018-nm tandem pumping is a rising technology that is deemed to produce fewer thermal loads for high-power YDFLs and to promise higher output power limits. The analysis here will treat these two examples.

For the 976-nm LD-pumped setup, other details are chosen according to the state-of-the-art design. The analysis then can be divided into the following steps.

(1) Suppose a pump optical power for the co-pumping setup. Use laser rate equations to predict the optical power distribution along the YDF. The result is shown as a pair of dashed lines in Fig. 9(a) (pump in blue and signal in red, both v.s. left vertical axis).

(2) Use Eq. 20/22 and Eq. 24 to obtain the heat source distribution $\dot{Q}_1(r,z)$ along the YDF. The result is shown as a solid line in Fig. 9(a) (v.s. right vertical axis).

(3) Use Eq. 27 to predict the operating temperature distribution of the YDF. The temperature distribution at the interface between silica cladding and polymer coating, $T(r_2)$, is shown as a solid line in Fig. 9(b). This parameter is concerning because the polymer coating can deteriorate and burn under high temperature before directly reaching the critical conditions for IFF; it can be a limiting factor that causes IFF after some time, depending on circumstances.

(4) Use the results from laser rate equations to calculate the total in-core optical powers that will be the responsible conditions for potential IFF. In calculating the total in-core optical power, the LD pumping light is considered uniform in the whole area of the core and cladding layers. The correlation with the core temperatures, $T(0)$, is shown as a solid line in Fig. 9(c). The solid lines of the same color among the 3 subplots of Fig. 9 are of the same pump optical powers (and thus the same output powers, as legend showing in Fig. 9(b)).

(5) Change for a new pump optical power, and repeat the above steps 1~4. This obtains a group of different lines in the 3 subplots of Fig. 9.

After the 976-nm LD-pumped setup is done, a typical 1018-nm tandem-pumped amplifier is also analyzed; the results are shown accordingly in Fig. 9(d~f). In calculating the total in-core optical power, the mode diameter of the 1018-nm tandem-pumping light needs to be considered. Here we use 30-μm-diameter as a typical value obtained from commercial solutions.

In the simulation, most of the values of laser signal power, i.e., those >3 kW in both 976-nm LD pumping and 1018-nm tandem pumping, can be unrealistic. It is not only because such signal powers are much higher than the to-date experimental records [2, 3, 41-45] that translate to ~1 kW in 10/130-μm YDFs. There are two more reasons. First, even if such high pump powers are available and are converted into signal powers, in-core optical powers >3 kW in 10/130-μm YDFs translate to optical power densities >4 GW/cm$^2$, which is about the LIDT in silica materials measured by pulsed lasers. Although >4 GW/cm$^2$ power density is still possible (such as [46]) (as the measuring conditions for LIDT have many differences with the operating conditions in high-power YDFLs), the upper limit of the power density is unlikely to be much higher. Second, it is expected that in-core signal powers >1 kW in 10/130-μm YDFs will induce SRS effect [2, 3, 28, 43-45] that evolves into strong pulses (hard to be quantitatively characterized) that finally triggers damage (manifesting as fiber fuse). That is, the operating status with SRS cannot be steady-state. Moreover, it is unclear whether other new limiting optical effects will emerge under such cases. This paper cannot discuss the relation between SRS or other limiting optical effects and fiber fuse, due to the lack of quantitative experimental results.

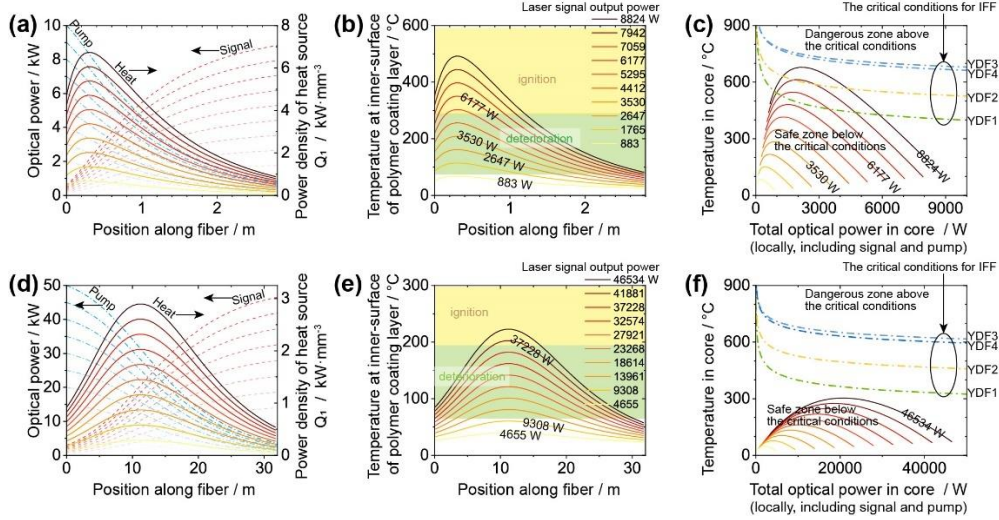

Fig. 9. Simulation results of the damage thresholds of continuous-wave 976-nm pumped YDFA (a~c) and 1018-nm pumped YDFA (d~f) in typical configurations. (a,d) Distribution of optical power (v.s. left vertical axis, pump power in blue dashed line, signal power in red dashed line), and the heat power density $\dot{Q}_1$ (v.s. right vertical axis, each line represents a setup of a different output power) along a presumed YDF. (b,e) Distribution of temperature of the inner surface of coating layer $T(r_2)$ along the YDF; yellow zone labels typical ignition points of the polymer coating layer. (c,f) Manifold showing the correlation between the in-core temperatures $T(0)$ and the in-core total optical powers (solid lines); the critical conditions for IFF of the 4 kinds of 10/130-μm YDFs are also shown (dashed lines) for comparison.

Nevertheless, the unrealistic cases of optical powers here are helping illustrate the potentials of the YDFs. For comparison, ranges of deterioration temperatures (starting from ~70°C) and ignition temperatures (starting from ~200°C) of typical polymer coating materials are shown as green and yellow blocks, respectively, in Fig. 9(b). Solid lines crossing the blocks there means that those cases take risks of polymer coating deterioration and burning, which is considered unsafe. The critical conditions for IFF in the 4 kinds of 10/130-μm YDFs are shown as dashed lines in Fig. 9(c). Solid lines crossing the dashed lines there means that the steady-state operating status of the setups (not considering the polymer coating burning) have reached the critical conditions for IFF, which is also unsafe.

As a result, in the 976-nm LD-pumped setup, the two situations impose different levels of risk. As shown in Fig. 9(b), ignition of polymer coating layers will happen at between ~3 kW to ~6 kW; deterioration of the polymer coating will happen at between ~1 kW to ~3 kW. If considering the deterioration as the limiting factor, the result well reflects the current development of 976-nm LD-pumped YDFLs based on 10/130-μm YDFs, where most studies reported output powers below 1 kW. As for Fig. 9(c), the experimentally revealed critical conditions for IFF are met at somewhere between ~6 kW to ~9 kW. It means that in the foreseeable future of experiment (powers <3 kW), spontaneous IFF won't be caused directly by the steady-state operating status of YDFLs itself, but may be caused by the deterioration and burning of the polymer coating layer, which lead to further temperature increases that finally cause IFF. In the 1018-nm tandem-pumped setup, the situation is alike, but the powers are much higher than those in the 976-nm pumped setup. As shown in Fig. 9(e), ignition of polymer coating will happen at >~35 kW, while deterioration will happen at >~10 kW. In fact, these limitations have not been met by to-date experimental capabilities. As shown in Fig. 9(f), the steady-state operating status of the 1018-nm tandem-pumped YDFL (not considering the limitation from polymer coating) cannot reach the critical conditions for IFF. Considering these results, spontaneous fiber fuses in the simulated 1018-nm tandem-pumped YDFL will require existence of some limiting optical effects, such as SRS, before they happen. The results also suggest that 1018-nm tandem pumping can produce considerably larger space for further increasing output powers of YDFLs than 976-nm LD pumping.

## 6. Conclusion

This paper presents majorly two results about fiber fuse in silica optical fibers. First, on the physical mechanisms of IFF, a method is proposed to simulate the critical conditions for IFF to prove the applicability of defect-induced absorption mechanism for explaining IFF. By simulating with suitable formation energies of defect $u_F$, the 3D solid-state heat transfer model with heat source from defect-induced absorption can produce simulation results of critical conditions for IFF that precisely match the experimental results. It is found that the previously experimentally revealed parameter $u_0$ in Eq. 2 is not directly the formation energy of defect $u_F$, but is the comprehensive effect from both defect chemical dynamics and heat dissipation, which can be explained by Eq. 30. The method can uniquely determine the defect formation energies $u_F$, which will have an important role in studying the physical mechanisms of fiber fuse in the future. Second, on the practical side, this paper offers a method to analyze the steady-state operating temperatures of gain fibers in continuous-wave fiber lasers. The general analytical solutions of the operating temperatures are deduced. The operating temperatures of both 976-nm LD-pumped and 1018-nm tandem-pumped YDF amplifiers using the 10/130-µm YDF are calculated for example. It is found that the steady-state operating status (the in-core temperatures as well as the in-core optical powers) of YDFLs of the state-of-the-art output power levels itself do not directly approach the critical conditions for IFF. But deterioration and burning of the polymer coating layers of the YDFs can be a limiting factor that imposes different power limits on the YDFLs. The results also suggest that 1018-nm tandem pumping provides wide space for further increasing the output powers of YDFLs. The power limits caused by fiber fuse will be far beyond the to-date experimental records, provided that limiting optical effects, including SRS, TMI, etc., can be well handled.

So far, there have not been more experimental measurements of the critical conditions for IFF in YDFs of other sizes, geometries or substance materials. Should more experimental data be there (with respective high monetary costs), in principle, the methods of this paper can be extended to treat fiber fuse problems in the respective kinds of YDFs. The same extension also applies to other kinds of optical fibers that have important and wide-spread applications, e.g., erbium-doped fibers (EDFs), thulium-doped fibers (TDFs) and holmium-doped fibers (HDFs). On the physical mechanism, the chemical cause of the defect formation energy $u_F$ still awaits studying, which will have a great influence on improving the resistance of the optical fibers against fiber fuse and achieving higher performance in their applications. On the application side, the results suggest that the polymer coating of the conventional double-cladding fibers is the major limiting factor that leads to fiber fuse. To alleviate this limitation, enhanced design of the cooling structures of optical fibers may have good effects, e.g., the recent advances in triple-cladding fibers [3, 4, 28] and metal-coating fibers [6, 7] are promising, which can alleviate or even remove the limitation of polymer coating deterioration and burning.

**Funding.** National Natural Science Foundation of China (62122040, 62075113, 61875103).

**Disclosures.** The authors declare no conflicts of interest.

**Data availability.** Data underlying the results presented in this paper are not publicly available at this time but may be obtained from the authors upon reasonable request.

1.